\documentstyle[fleqn,twoside,gc]{article}

\def\be{\begin{equation}}
\def\ee{\end{equation}}
\def\bea{\begin{eqnarray}}
\def\eea{\end{eqnarray}}
\def\beaa{\begin{eqnarray*}}
\def\eeaa{\end{eqnarray*}}
\def\nn{\nonumber \\}
\def\e{{\rm e}}
\def\CC{{\left(4\epsilon - 1\right)^2 \over 4\kappa_g^4}}
\def\DD{\left(2\epsilon{\tilde{\mu}} \left(4\epsilon  - 1\right)\right)}

\heads{Shin'ichi Nojiri}
      {Higher Derivative Gravities and Negative Entropy}

\begin{document}
\twocolumn[
\Arthead{6}{2000}{4 (24)}{1}{10}

\Title{Higher Derivative Gravities  \yy
       and Negative Entropy}


   \Author{Shin'ichi Nojiri\foom 1}   
          {Department of Applied Physics, National Defence Academy, 
Hashirimizu Yokosuka 239-8686, JAPAN}              

\Abstract
{We investigate the black hole solutions in the $R^2$-gravity, where the
action
contains the square of the curvature. In case that the action does not contain
the square of the Riemann tensor and in case that the $R^2$-terms are the
Gauss-Bonnet (GB) combination, we find exact solutions. We investigate the
thermodynamics of these theories and find the Hawking-Page like phase
transition, which is the phase transition between the black hole (BH)
spacetime and the pure anti-deSitter (AdS) spacetime. From the viewpoint of
the
AdS/CFT correspondence, such a phase transition may correspond to thermal
transition  of dual CFT. 

An interesting feature of $R^2$-gravity is the possibility of the negative (or
zero) dS (or AdS) BH entropy, which depends on the parameters of the
$R^2$-terms. We speculate that the appearence of negative entropy may indicate
a new type instability where a transition between dS (AdS) BH with negative
entropy and AdS (dS) BH with positive entropy occurs.

We also apply the GB gravity to the brane cosmology, where the brane moves in
the bulk AdS BH spacetime. By investigating the FRW-like equation, which
describes the motion of the brane, we find the behavior of the matter on the
brane. When the radius of the brane is large, the matter fields behave as CFT
but when the radius is small, the brane universe behaves as the universe with
dust or curvature dominant universe, depending on the parameters.}



]  
\email 1 {snojiri@yukawa.kyoto-u.ac.jp, nojiri@cc.nda.ac.jp}

\section{Introduction\label{Sec1}}

The Einstein-Hilbert action includes only scalar curvature
\be
\label{I}
S_{\rm Einstein}=\int d^{d+1}x\sqrt{-g}\left\{{1\over \kappa^2}R-\Lambda 
\right\}+ S_{\rm matter}\ , 
\ee
Here $S_{\rm matter}$ is the action of matter fields. On the other hand the
action of $R^2$-gravity includes the square of the curvatures
\bea
\label{II}
S&=&\int d^{d+1}x\sqrt{-g}\left\{aR^2+bR_{\mu\nu}R^{\mu\nu}\right.\nn
&&\left.+cR_{\mu\nu\xi\sigma}R^{\mu\nu\xi\sigma}\right\}+S_{\rm Einstein}\ .
\eea
One of the reasons why we consider the $R^2$-gravity is because the limit of
$\alpha'\rightarrow 0$ limit in string theory corresponds the Einstein gravity
but if we include the first order correction with respect to $\alpha'$, we
obtain $R^2$-gravity. From the viewpoint of the AdS/CFT correspondence
\cite{AdS} (or recent dS/CFT correspondence\cite{strominger,ds1,noplb}),
$\alpha'$ correction corresponds to $1/N$ corrections in the field theory
side. The equations of motion of the $R^2$-gravity, which correspond
to the Einstein equation has the following form:
\bea
\label{III}
&&\left.0={1\over 2}g^{\mu\nu}\right.(aR^2+bR_{\mu\nu}R^{\mu\nu}
+cR_{\mu\nu\xi\sigma}R^{\mu\nu\xi\sigma}\nn
&&\left.+{1\over\kappa^2}R-\Lambda\right)+a\left(-2RR^{\mu\nu}\right.\nn
&&\left.+\nabla^\mu\nabla^\nu R+\nabla^\nu\nabla^\mu R-2g^{\mu\nu}\nabla_\rho
\nabla^\rho R\right)\nn
&&+b\left(-2R^\mu_{\ \rho}R^{\nu\rho}+\nabla_\rho\nabla^\mu R^{\rho\nu}
+\nabla_\rho\nabla^\nu R^{\rho\mu}\right.\nn
&&\left.-\Box R^{\mu\nu}-g^{\mu\nu}\nabla^\rho\nabla^\sigma R_{\rho\sigma}
\right)\nn
&&+c\left(-2R^{\mu\rho\sigma\tau}R^\nu_{\ \rho\sigma\tau}
 -2\nabla_\rho\nabla_\sigma R^{\mu\rho\nu\sigma}\right.\nn
&&\left.-2\nabla_\rho\nabla_\sigma R^{\nu\rho\mu\sigma}
\right)-{1\over\kappa^2}R^{\mu\nu}+T^{\mu\nu} \ .
\eea
Here $T^{\mu\nu}$ is energy-momentum tensor of matter: $T^{\mu\nu}={1\over
\sqrt{-g}}{\delta S_{\rm matter} \over \delta g_{\mu\nu}}$. The above equation
(\ref{III}) is very complicated but there are several solvable cases. One is
$c=0$ case, where the Riemann tensor $R_{\nu\rho\mu\sigma}$ does not appear in
the equation of motion, then the solutions can be constructed from the
solutions of the Einstein equation. The other is the Gauss-Bonnet (GB)
invariant combination ($a=c$, $b=-4c$) case. Of course, the GB term is trivial
(topological) in 4 dimensions but non-trivial when
the dimensions of the spacetime are more than 4.

We find the black hole (BH) solutions for the $c=0$ case in Section \ref{Sec2}
and for the Gauss-Bonnet combination case in Section \ref{Sec3}. By using the
obtained solution, we investigate the thermodynamics of the $R^2$-gravity in
Section \ref{Sec4} and we find the Hawking-Page like phase transition, which
is the phase transition between the black hole spacetime and the pure AdS
spacetime. From the viewpoint of the AdS/CFT correspondence, such a phase
transition may be that of the field theory when we include $1/N$ corrections.
An interesting feature of higher derivative gravity is the possibility of the
negative (or zero) dS (or AdS) BH entropy which  depends on the parameters of
higher derivative terms. In Section \ref{Sec5}, we investigate the problem of
the negative entropy and we speculate that the appearence of negative entropy
may indicate a new type instability where a transition between dS (AdS) BH
with negative entropy and AdS (dS) BH with positive entropy occurs. In Section
\ref{Sec5}, we apply the GB gravity to the brane cosmology, where the brane
moves in the bulk AdS black hole spacetime. By investigating the FRW-like
equation, which describes the motion of the brane, we find the behavior of the
matter fields on the brane. When the radius of the brane is large, the matter
fields behave as CFT but when the radius is small, the brane universe behaves
as the universe with dust or curvature dominant universe, depending on the
parameters. The last section is devoted to the summary.

This report is mainly based on \cite{NOO1,NOO2,CNO,NO1,NO2,LNO,NOO3}.

\section{Solutions of $c=0$ case\label{Sec2}}

We first consider the solutions of $c=0$ case. The $D=d+1$ dimensional
Einstein equation without matter is, of course,  given by
\be
\label{VI}
R_{\mu\nu}-{1\over 2}g_{\mu\nu}R=-{\kappa^2\over 2}\Lambda g_{\mu\nu}\ .
\ee
Then by multiplying $g^{\mu\nu}$, we find the scalar curvature is constant
${d-1\over 2}R={d+1\over 2}\kappa^2\Lambda$ and the Ricci tensor is
proportional to the metric $R_{\mu\nu}={d\over 2(d-1)}\kappa^2\Lambda
g_{\mu\nu}$ and therefore covariantly constant $\nabla_\rho R_{\mu\nu}$ due to
the condition of the metricity $\nabla_\rho g_{\mu\nu}=0$. We now apply this
results to the $R^2$-gravity with $c=0$. 

By assuming
\be
\label{VII}
R=-{d(d+1)\over l^2}\ ,\quad R_{\mu\nu}=-{d\over l^2}g_{\mu\nu}\ ,
\ee
and substituting these expressions of the curvatures
to Eq.(\ref{III}) with $c=0$, we obtain
\be
\label{VIII}
0={d^2(d+1)(d-3)a\over l^4}+{d^2(d-3)b\over l^4}-{d(d-1)\over\kappa^2
l^2}-\Lambda\ ,
\ee
from which $l^2$ can be determined. Then if we choose ${1\over l^2}=-{d\over
2(d-1)}\kappa^2\Lambda$ in the Einstein equation (\ref{VI}), we can construct
a solution of $R^2$-gravity (with $c=0$) from the solution of the vacuum
Einstein equation.

\section{Solutions in the Gauss-Bonnet combination case \label{Sec3}}

In this section, we consider the solutions in the Gauss-Bonnet combination.

By using Bianchi identity
\be
\label{IX}
\nabla_\mu R^\lambda_{\ \sigma\nu\rho}+\nabla_\rho R^\lambda_{\ \sigma\mu\nu}
+\nabla_\nu R^\lambda_{\ \sigma\rho\mu}=0\ ,
\ee
we obtain
\bea
\label{X}
&&\nabla_\rho\nabla_\sigma R^{\mu\rho\nu\sigma}\nn
&&=\Box R^{\mu\nu}-{1 \over 2}\nabla^\mu \nabla^\nu R
+R^{\mu\rho\nu\sigma}R_{\rho\sigma}-R^\mu_{\ \rho}R^{\nu\rho}\ ,\nn
&&\nabla_\rho\nabla^\mu R^{\rho\nu}+\nabla_\rho\nabla^\nu R^{\rho\mu}\nn
&&={1\over 2}\left(\nabla^\mu \nabla^\nu R+\nabla^\nu\nabla^\mu R\right) \nn
&&\ -2R^{\mu\rho\nu\sigma} R_{\rho\sigma}+2R^\mu_{\ \rho}R^{\nu\rho}\ ,\nn
&&\nabla_\rho\nabla_\sigma R^{\rho\sigma}={1\over 2}\Box R \ .
\eea
By using the above equations (\ref{X}), we can rewrite
the equations of motion (\ref{III}) as follows:
\bea
\label{XI}
&&\left.0={1\over 2}g^{\mu\nu}\right(aR^2+bR_{\mu\nu}R^{\mu\nu}
+cR_{\mu\nu\xi\sigma}R^{\mu\nu\xi\sigma}\nn
&&\left.+{1\over\kappa^2}R-\Lambda\right)\nn
&&+a\left(-2RR^{\mu\nu}+\nabla^\mu\nabla^\nu R+\nabla^\nu\nabla^\mu R -2
g^{\mu\nu}\Box R\right) \nn
&&+b\left\{{1\over 2}\left(\nabla^\mu\nabla^\nu R+\nabla^\nu\nabla^\mu
R\right)-2 R^{\mu\rho\nu\sigma}R_{\rho\sigma}\right.\nn
&&\left.-\Box R^{\mu\nu}-{1 \over 2}g^{\mu\nu}\Box R\right\}\nn
&&+c\left(-2R^{\mu\rho\sigma\tau}R^\nu_{\ \rho\sigma\tau}-4\Box R^{\mu\nu}
+\nabla^\mu\nabla^\nu R\right.\nn
&&\left.+\nabla^\nu \nabla^\mu R-4R^{\mu\rho\nu\sigma}R_{\rho\sigma}
+4R^\mu_{\ \rho}R^{\nu\rho}\right)\nn
&&-{1\over\kappa^2}R^{\mu\nu}-T_{\rm matter}^{\mu\nu}\ .
\eea
When the coefficients $a$, $b$, $c$ are Gauss-Bonnet combination: $a=c$ and
$b=-4c$, one finds
\bea
\label{XIII}
&&\left.0={1\over 2}g^{\mu\nu}\right\{c\left(R^2-4R_{\mu\nu}R^{\mu\nu}
+R_{\mu\nu\xi\sigma}R^{\mu\nu\xi\sigma}\right)\nn
&&\left.+{1\over\kappa^2}R-\Lambda\right\}
+c\left(-2RR^{\mu\nu}+4 R^\mu_{\ \rho}R^{\nu\rho}\right.\nn
&&\left.+4R^{\mu\rho\nu\sigma}R_{\rho\sigma}-2R^{\mu\rho\sigma\tau}
R^\nu_{\ \rho\sigma\tau}\right)\nn
&&-{1\over \kappa^2}R^{\mu\nu}+T^{\mu\nu} \ .
\eea
Since the above equation does not contain the derivative of the curvarutre,
the equation contains only second order derivative of the metric tensor. As a
classical theory, if we impose the initial conditions for metric tesor and its
first derivative with respect to time, the metric tensor can be determined
uniquely. 

We now find a solution. We now assume the metric in the following form:
\be
\label{XIV}
ds^2=-\e^{2\nu(r)}dt^2+\e^{2\lambda(r)}dr^2
+r^2\sum_{i,j=1}^{d-1}\tilde g_{ij}dx^idx^j\ .
\ee
Here $\tilde g_{ij}$ : the metric tensor of the Einstein manifold defined by
$\tilde R_{ij}=k g_{ij}$. Here $\tilde R_{ij}$ is the Ricci tensor given by
$\tilde g_{ij}$ and $k$ is a constant. For example, $k=d-2$ corresponds to the
$d-1$-dimensional unit sphere, $k=-(d-2)$ to $d-1$-dimensional unit
hyperboloid and $k=0$ to a flat surface.

We consider the electromagnetic fields as a matter:
\bea
\label{XV}
S_{\rm matter}&=&-{1\over 4g^2}\int d^{d+1}x \sqrt{-g}
g^{\mu\nu}g^{\rho\sigma}F_{\mu\rho}F_{\nu\sigma}\ ,\\
F_{\mu\nu}&=&\partial_\mu A_\nu-\partial_\nu A_\mu\ .
\eea
Here $A_\mu$ is a vector potential (gauge field) and $g$ is the gauge
coupling.
If $F_{tr}=-F_{rt}$ only depend on $r$ and the other components
of $F_{\mu\nu}$ vanish, one finds
\be
\label{XVI}
F_{tr}=\e^{\nu+\lambda}r^{1-d}Q
\ee
Here $Q$ is a constant corresponding to the electric charge.
Then  the solution, which was first found in \cite{Wil}, appears
\bea
\label{XVII}
&&\e^{2\nu}=\e^{-2\lambda}\nn
&&={1\over 2c}\left[{2ck \over d-2}+{r^2 \over \kappa^2 (d-2)(d-3)}\right.\nn
&&\pm \left\{ {r^4 \over \kappa^4 (d-2)^2(d-3)^2}  \right. \nn
&&+{4c\Lambda r^4\over d(d-1)(d-2)(d-3)}-{2cQ^2r^{6-2d}\over g^2
(d-1)(d-3)}\nn
&&\left.\left.-{4cCr^{4-d}\over (d-1)(d-2)(d-3)}\right\}^{1\over 2}\right]\ .
\eea
The radius $r=r_H$ of the event horizon, where $\e^{2\nu}=0$ or
$\e^{-2\lambda}=0$, is given by
\bea
\label{XVIII}
0&=&{(d-1)(d-3)\over d-2}ck^2r_H^{d-4}+{(d-1)k\over (d-2)\kappa^2}r_H^{d-2}\nn
&&-{\Lambda\over d}r_H^d+{(d-2)Q^2\over 2g^2}r_H^{2-d}+ C\ ,
\eea
The Haking temperature $T_H$ is also given by
\bea
\label{XIX}
&&4\pi T_H=\left.\left(\e^{2\nu}\right)'\right|_{r=r_H}\nn
&&={1\over 2}\left({2ck\over
d-2}+{r_H^2\over\kappa^2(d-2)(d-3)}\right)^{-1}\nn
&&\times\left[{4kr_H\over\kappa^2(d-2)^2(d-3)}\right.\nn
&&-{8\Lambda r_H^3\over d(d-1)(d-2)(d-3)}-{2Q^2r_H^{5-2d}\over (d-1)g^2}\nn
&&\left.-{2(d-4)Cr_H^{3-d}\over (d-1)(d-2)(d-3)}\right]\ .
\eea
Especially for $d=4$, we have
\bea
\label{XX}
&&\e^{2\nu}=\e^{-2\lambda}\nn
&&={1\over 2c}\left[ck+{r^2\over 2\kappa^2}\right.\nn
&&\left.\pm\left\{{r^4\over 4\kappa^4}+{c\Lambda r^4 \over 6}-{2cQ^2
\over 3g^2r^2 }-{2c C\over3}\right\}^{1\over 2}\right]\ .
\eea
When $r$ is large, the obtained solution behaves as
\bea
\label{XXI}
&&\e^{2\nu}=\e^{-2\lambda}\nn
&&={1\over 2c}\left[{r^2\over 2\kappa^2 }\left(1\pm
\sqrt{1+{2c\Lambda\kappa^4\over 3}}\right)+ck\right.\nn
&&\mp{2\kappa^2cC\over3r^2\sqrt{1+{2c\Lambda\kappa^4\over 3}}}\nn
&&\left.\mp{2cQ^2\over 3g^2r^4\sqrt{1+{2c\Lambda\kappa^4\over 3}}}
+{\cal O}\left(r^{-4}\right)\right]\ .
\eea
We compare the above behavior with the
Reissner-Nordstr\o m-AdS solution in the Einstein gravity:
\be
\label{XXII}
\e^{2\nu_{\rm RNAdS}}=\e^{-2\lambda_{\rm RNAdS}}
={r^2\over l^2}+{k\over 2}-{\mu\over r^2}+{q^2\over r^4}\ ,
\ee
and we find
\bea
\label{XXIII}
&&{1\over l^2}={1\over 4c\kappa^2}\left(1\pm\sqrt{1+{2c\Lambda\kappa^4\over
3}}\right)\ ,\nn
&&\mu=\pm{\kappa^2C\over 3\sqrt{1+{2c\Lambda\kappa^4\over 3}}}\ ,\nn
&&q^2=\mp{Q^2\over 3g^2\sqrt{1+{2c\Lambda\kappa^4\over 3}}}\ ,
\eea
that is
\bea
\label{XXIV}
&&\Lambda =-{12\over\kappa^2l^2}+{24c\over l^4}\ ,\quad
C=\mu\left({12c\over l^2}-{3\over \kappa^2}\right)\ ,\nn
&&{Q^2\over g^2}=3q^2\left(1-{4c\kappa^2\over l^2}\right)\ . 
\eea

\section{Hawking-Page phase transition in $R^2$-gravity\label{Sec4}}

Since the black hole has temperature, one can consider the thermodynamics
of the
spacetime. The theory with finite temperature for an action $S(\phi)$ is given
by Wick-rotating the time coordinate ($t\rightarrow it$) and imposing the
periodic boundary condition for $t$ with the period $\beta$: $Z(\beta)=\int[d
\phi]\e^{S(\phi)}$. The period $\beta$ can be regarded as the inverse of the
temperature $T$: $\beta ={1\over k_BT}$. Here $k_B$ is the Boltzmann constant
and we put $k_B=1$ in the following. Then $Z(\beta)$ can be regarded as the
thermodynamical partition function. In the WKB approximation, by substituting
the classical solution  $\phi=\phi_{\rm class}$ into the action, we obtain the
free energy $F$ by $F=-T\ln Z(\beta)=-TS\left(\phi_{\rm class}\right)$. In the
following, we concentrate on the $D=d+1=5$ dimensional asymptotically AdS
spacetime. 

Before going to the $R^2$-gravity, we consider the case of Einstein gravity
without matter. When the spacetime metric is given by
\be
\label{XXV}
ds^2=-\e^{2\nu(r)}dt^2+\e^{-2\nu(r)}dr^2+r^2\sum_{i,j=1}^3\tilde g_{ij}dx^i
dx^j\ ,
\ee
we find 
\bea
\label{XXVI}
&&F=-T\ln Z(\beta)=-TS_{\rm Einstein}\nn
&&S_{\rm Einstein}=\int
d^5x\sqrt{g}\left\{{1\over\kappa^2}R-\Lambda\right\}\nn
&&=-{8V_3\over Tl^2\kappa^2}\int_{r_H}^\infty dr\,r^3\ ,\nn
&&\Lambda=-{12\over l^2\kappa^2}\ ,\quad V_3=\int d^3x\sqrt{\tilde g}\ .
\eea
Then  $S_{\rm Einstein}$ diverges.  The regularization of the
divergence is given by cutting off the integral at a large radius $r=
r_{\rm max}$ and subtracting the solution of pure but finite temperature AdS:
\bea
\label{XXVII}
&&S_{\rm reg}=-{8V_3\over T\kappa^2l^2}\left\{\int^{r_{\rm max}}_{r_H}dr
r^3\right.\nn
&&\left.-\e^{\nu(r=r_{max})-\rho(r=r_{\rm max};\mu =0)}
\int^{r_{\rm max}}_{0}drr^{3}\right\} 
\eea
The factor $\e^{\nu(r=r_{\rm max})-\nu(r=r_{\rm max};\mu = 0)}$ is chosen so
that the proper length of the circle which corresponds to the period $\beta={1
\over T_H}$ in the Euclidean time at $r=r_{\rm max}$ coincides with each other
in the two solutions. Especially for $k=2$, we find in the limit of
$r_{\rm max}\to\infty$
\be
\label{XXVIII}
F=-{V_3\over\kappa^2}r_H^2\left({r_H^2\over l^2}-1\right)\ .
\ee
When ${r_H^2\over l^2}>1$, the free energy becomes negative: $F<0$, which
tells
that black hole spacetime is more stable than the pure AdS spacetime. That is,
large black holes are stable but small ones are instable and there is a
critical point at ${r_{H}^{2} \over l^{2}}=1$, which is the famous
Hawking-Page
phase transition \cite{HP}.

In $c=0$ $R^2$-gravity case, we find
\be
\label{XXIX}
F=-{V_3\over 8}r_H^2\left({r_H^2\over l^2}-{k\over 2}\right)\left({8\over
\kappa^2}-{320a\over l^2}-{64b\over l^2}\right)\; ,
\ee
which is similar to that in the Einstein gravity case
(\ref{XXVIII}) but there is a critical point (surface) at
\be
\label{XXX}
{8\over\kappa^2}-{320a\over l^2}-{64b\over l^2}=0\ .
\ee
Therefore when ${8\over\kappa^2}-{320a \over l^2}-{64b\over l^2}<0$, small
black hole is stable.

In case of the Gauss-Bonnet-Maxwell black hole case, we find
\bea
\label{XXXI}
F&=&-V_3\left\{-3\left({2c\over l^4}-{1\over
\kappa^2l^2}\right)r_H^4\right.\nn
&&\left.-2\pi T_Hr_H\left({r_H^2\over\kappa^2}-12c\right)\right\}\ .
\eea
Since we have two independent parameters $\mu$ and $Q^2$, $r_H$ can be
independent of $T_H$. Then  the critical line, where $F=0$,
is given by
\be
\label{XXXII}
T_H=T_c \equiv{3\over 2\pi}\left({r_H^2\over\kappa^2}-6ck\right)^{-1}\left({1
\over\kappa^2 l^2}-{2c\over l^4}\right)r_H^3\ .
\ee
If $T_H>T_c$ and $r_H^2 >12c\kappa$ (or, $T_H<T_c$ and $r_H^2<12c\kappa$),
pure
anti-de Sitter spacetime is more stable than the black hole spacetime. On the
other hand, if $T_H<T_c$ and $r_H^2>12c\kappa$ (or, $T_H>T_c$ and $r_H^2<12c
\kappa$), vice versa. In order to simplify the expressions, we (re)define the
following parameters and the variables:
\be
\label{XXXIII}
\epsilon\equiv{c\kappa^2\over l^2}\ ,\quad
r_H\rightarrow lr_H\ ,\quad T_H\rightarrow{T_H\over l}\ .
\ee
Then the critical temperature $T_c$ is expressed as
\be
\label{XXXIV}
T_c={3\left(1-2\epsilon\right)r_H^3\over 2\pi\left(r_H^2-12\epsilon\right)}\ .
\ee
>From the above expression (\ref{XXXIV}), one gets:
\begin{itemize}
\item When $\epsilon<0$ $\left(<{1 \over 2}\right)$, $T_c$ is always positive.
Then if $T_H>T_c$ $\left(T_H<T_c\right)$, the pure AdS (black hole) spacetime
is more stable than black hole (pure AdS) spacetime.
\item When $0<\epsilon<{1 \over 2}$, the critical temperature $T_c$ is
positive
when $r_H^2>12\epsilon$ and there is a critical line, where if $T_H>T_c$ $
\left(T_H<T_c\right)$, the pure AdS (black hole) spacetime is more stable than
the black hole (pure AdS) spacetime. When $r_H^2<12\epsilon $, $T_c$ is
negative, then the black hole spacetime is always stable. 
\item When $\epsilon>{1 \over 2}$, if $r_H^2>12\epsilon$, $T_c$ is negative
and the pure AdS spacetime is more stable than the black hole spacetime. If
$r_H^2<12\epsilon$, $T_c$ is positive and if $T_H>T_c$ $\left(T_H<T_c\right)$,
the black hole (pure AdS) spacetime is more stable than the pure AdS
(black hole) spacetime.
\end{itemize}
The conceptual (not-exact) Hawking-Page phase diagrams are given by Figures
\ref{Fig1a}, \ref{Fig1b} and \ref{Fig1c}.

\unitlength=0.5mm
\begin{center}
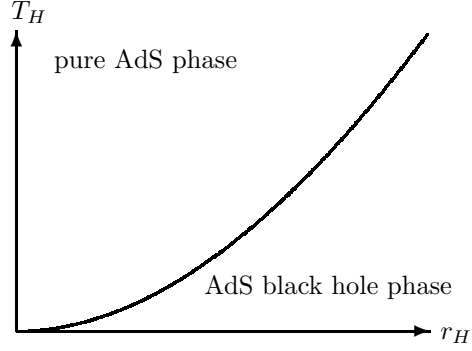
\begin{figure}
\begin{picture}(150,150)
\thicklines
\put(8,133){$T_H$}
\put(122,48){$r_H$}
\put(10,50){\vector(1,0){110}}
\put(10,50){\vector(0,1){80}}
\qbezier[400](10,50)(60,50)(119,129)
\put(20,120){pure AdS phase}
\put(60,60){AdS black hole phase}
\end{picture}

\vskip -2cm

\caption{The phase diagrams when $\epsilon<0$. \label{Fig1a}}
\end{figure}

\begin{figure}
\begin{picture}(150,150)
\thicklines
\put(8,133){$T_H$}
\put(122,48){$r_H$}
\put(10,50){\vector(1,0){110}}
\put(10,50){\vector(0,1){80}}
\qbezier[300](51,129)(51,30)(119,129)
\put(30,60){AdS black hole phase}
\put(55,120){pure AdS}
\put(65,110){phase}
\put(35,35){$r_H^2=12\epsilon $}
\thinlines
\put(50,50){\line(0,1){80}}
\put(50,43){\vector(0,1){7}}
\end{picture}

\vskip -1.5 cm

\caption{The phase diagrams when $0<\epsilon<{1\over 2}$. \label{Fig1b}}
\end{figure}

\begin{figure}
\begin{picture}(170,150)
\thicklines
\put(38,133){$T_H$}
\put(152,48){$r_H$}
\put(40,50){\vector(1,0){110}}
\put(40,50){\vector(0,1){80}}
\qbezier[200](40,50)(79,80)(79,129)
\put(75,60){pure AdS phase}
\put(48,120){AdS}
\put(46,110){black}
\put(48,100){hole}
\put(46,90){phase}
\put(75,35){$r_H^2=12\epsilon $}
\thinlines
\put(80,50){\line(0,1){80}}
\put(80,43){\vector(0,1){7}}
\end{picture}
\vskip -1.5 cm
\caption{The phase diagrams when $\epsilon>{1\over 2}$. \label{Fig1c}}
\end{figure}
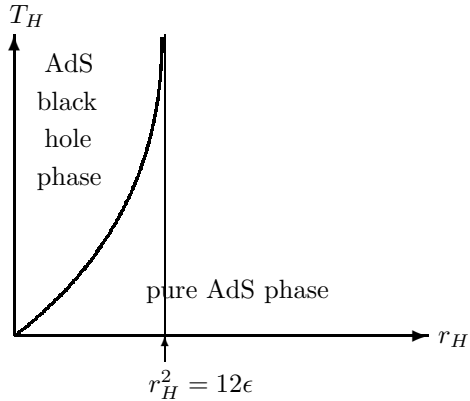

\end{center}

\section{Negative Entropy?\label{Sec5}}

In this section, based on the entropy, we investigate the relation between
Schwarzschild-de Sitter black hole and Schwarzschild-{\it anti} de Sitter
black hole.

In the spacetime with black hole, we put a boundary, which is a surface with
constant $r$. On the boundary, one takes an action, which makes the
variational
principle well-defined (like Gibbons-Hawking term) and further makes the total
action finite. Then we can define an energy-momentum tensor on the boundary
and we can determine the mass $M$ ($d=4$):
\bea
\label{XXXV}
E&=&M \nn
&=&{3l^2\over 16}V_3\left({1\over\kappa^2}-{40a\over l^2}-{8b\over l^2}-{4c
\over l^2}\right)\nn
&&\times\left(k^2+{16\mu\over l^2}\right)\ .
\eea
 $M$ is thermodynamical energy $E$ \cite{BK}.  

By using the thermodynamical relation $d{\cal S}={dE \over T}$, in $c=0$ case,
one arrives at
\be
\label{XXXVI}
{\cal S}=\int{dE\over T_H}={V_3\pi r_H^3\over 2}\left({8\over\kappa^2}-{320a
\over l^2}-{64b\over l^2}\right)+ {\cal S}_0\ .
\ee
Here $S_0$ is a constant of the integration.
Then there might appear a natural question, that is, when
\be
\label{XXXVII}
{8\over \kappa^2}-{320a\over l^2}-{64b \over l^2}<0\ ,
\ee
is entropy negative? Even if $S_0\neq 0$, the entropy becomes negative for the
large (large $r_H$) black hole. We should note that $l^2$ ($d=4$) is given by
solving Eq.(\ref{VIII}) with $d=4$, which can be rewritten as
\bea
\label{XXXVIII}
0&=&l^4+{12l^2\over\kappa^2\Lambda}-{80a+16b\over\Lambda}\nn
&=&\left(l^2+{5\over\kappa^2\Lambda}\right)^2-{36\over\kappa^4\Lambda^2}-{80a
+16b\over\Lambda}\ .
\eea
Then if
\be
\label{XXXIX}
{36\over\kappa^4\Lambda^2}+{80a+16b\over\Lambda}\geq 0\ ,
\ee
there are real solution of $l^2$:
\be
\label{IVX}
{1\over l^2}={{6\over\kappa^2}\pm\sqrt{{36\over\kappa^4}
+\Lambda(80a+16b)}\over80a+16b}\ .
\ee
Then from Figure \ref{Fig2}, we find
\begin{enumerate}
\item When $80a+16b>0$ and $\Lambda>0$, or $80a+16b<0$ and $\Lambda<0$, one of
the solutions of $l^2$ is positive but another is negative. Then there are two
kind of solution, one is asymptotically anti-de Sitter and another is
asymptotically de Sitter.
\item When $80a+16b>0$ and $\Lambda<0$, $l^2$ is always positive. Both of two
solutions express the asymptotically anti-de Sitter spacetime.
\item When $80a+16b<0$ and $\Lambda>0$, $l^2<0$. Both of two
solutions correspond to asymptotically de Sitter spacetime.
\end{enumerate}

\unitlength=0.3mm

\begin{center}

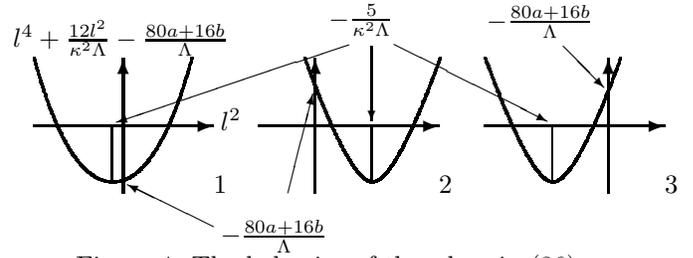
\begin{figure}
\begin{picture}(300,80)
\thicklines
\put(0,75){$l^4+{12l^2\over\kappa^2\Lambda}-{80a+16b\over\Lambda}$}
\put(92,38){$l^2$}
\put(10,40){\vector(1,0){80}}
\put(50,10){\vector(0,1){60}}
\put(90,10){1}
\qbezier[200](10,70)(25,15)(45,15)
\qbezier[200](80,70)(65,15)(45,15)
\put(110,40){\vector(1,0){80}}
\put(135,10){\vector(0,1){60}}
\put(190,10){2}
\qbezier[200](130,70)(150,15)(160,15)
\qbezier[200](190,70)(170,15)(160,15)
\put(210,40){\vector(1,0){80}}
\put(265,10){\vector(0,1){60}}
\put(290,10){3}
\qbezier[200](210,70)(230,15)(240,15)
\qbezier[200](270,70)(250,15)(240,15)
\thinlines
\put(45,15){\line(0,1){25}}
\put(160,15){\line(0,1){25}}
\put(240,15){\line(0,1){25}}
\put(150,76){\vector(-3,-1){103}}
\put(160,75){\vector(0,-1){33}}
\put(170,76){\vector(2,-1){68}}
\put(140,84){$-{5\over\kappa^2\Lambda}$}
\put(90,-5){\vector(-2,1){38}}
\put(123,10){\vector(1,4){11}}
\put(245,75){\vector(1,-1){18}}
\put(92,-12){$-{80a+16b\over\Lambda}$}
\put(210,83){$-{80a+16b\over\Lambda}$}
\end{picture}
\caption{The behavior of the r.h.s. in (\ref{XXXVIII}).\label{Fig2}}
\end{figure}

\end{center}

The expression of the entropy (\ref{XXXVI}) can be applied even for
asymptotically de Sitter spacetime by putting $l^2=-l_{\rm dS}^2<0$. By the
explicit expression of $l^2$ via $\Lambda$, $a$, $b$, and $\kappa^2$, we
obtain
\be
\label{IVXI}
{\cal S}={V_3\pi r_H^3\over 2}\left(-{16\over\kappa^2}\mp\sqrt{{36
\over\kappa^4}+\Lambda(80a+16b)}\right)\ .
\ee
If we assume the entropy should be positive, we need to choose the lower sign
$+$ in $\pm$ since the entropy always becomes negative if we choose the upper
sign $-$. Even if one chooses the lower sign $+$, the entropy is not always
positive but it becomes positive when
\be
\label{IVXII}
{20\over\kappa^4\Lambda^2}+{80a+16b\over\Lambda}\geq 0\ .
\ee
Then we obtain the diagram in Figure \ref{Fig3}.

\unitlength=0.28mm

\begin{center}

\begin{figure}
\begin{picture}(300,200)
\thicklines
\small
\put(140,193){$80a+16b$}
\put(292,98){$\Lambda$}
\put(10,100){\vector(1,0){280}}
\put(150,10){\vector(0,1){180}}
\qbezier[500](10,110)(135,115)(140,190)
\qbezier[500](290,90)(165,85)(160,10)
\qbezier[500](10,120)(120,120)(130,190)
\qbezier[500](290,80)(180,70)(170,10)
\put(182,178){$\Lambda(80a+16b)+{20 \over \kappa^4}=0$}
\put(0,25){$\Lambda(80a+16b)+{36 \over \kappa^4}=0$}
\put(160,120){dS region}
\put(160,80){dS region}
\put(70,105){AdS region}
\put(55,70){AdS region}
\put(157,158){negative entropy}
\put(172,148){region}
\put(20,170){forbidden region}
\put(200,35){forbidden region}
\thinlines
\put(139,180){\line(1,0){41}}
\put(280,89){\line(-1,2){43}}
\put(20,120){\line(1,-2){43}}
\put(162,28){\line(-1,0){40}}
\put(155,160){\line(-1,0){30}}
\put(220,150){\line(1,-2){35}}
\end{picture}
\caption{The phase diagram of the spacetime with respect to the parameters.
\label{Fig3}}
\end{figure}
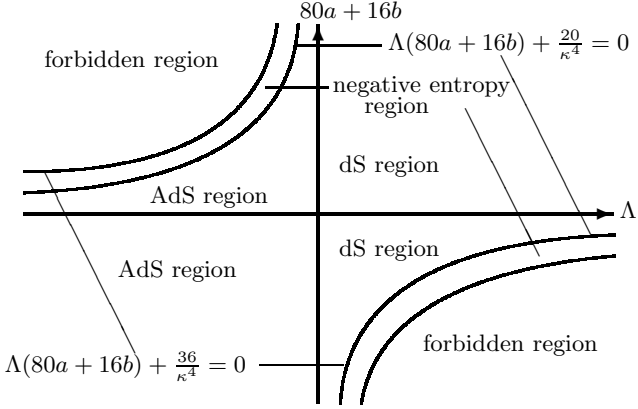

\end{center}

In the Gauss-Bonnet case, we obtain
\be
\label{IVXIIb}
E=M={3l^2\over 16}V_3\left({1\over\kappa^2}-{12c\over l^2}
\right)\left(k^2+{16\mu\over l^2}\right)
\ee
and therefore
\be
\label{IVXIII}
{\cal S}={V_3\over\kappa^2}\left({1-12\epsilon\over 1-4\epsilon}\right)\left(
4\pi r_H^3+24\epsilon k\pi r_H\right)+{\cal S}_0\ ,
\ee
Here $\epsilon\equiv {c\kappa^2 \over l^2}$ and
the length parameter $l^2$ is given by
\be
\label{IVXIV}
{1\over l^2}={1\over 4c\kappa^2}\left(1\pm\sqrt{1+{2c\Lambda\kappa^4\over 3}}
\right)\ .
\ee
Then in order that $l^2$ is a real number, we have
\be
\label{IVXIVb}
c\Lambda\geq-{3 \over 2\kappa^4}\ .
\ee
Then we find
\begin{enumerate}
\item When $c\Lambda>0$, there exist two kind of solutions, one expresses
asymptotically AdS spacetime and another expresses the asymptotically dS
spacetime.
\item When $c>0$ and $\Lambda<0$, both of two solutions expresses
asymptotically AdS spacetime.
\item When $c<0$ and $\Lambda>0$, both of solutions correspond to
asymptotically dS spacetime.
\end{enumerate}
Substituting the expression of $l^2$, we obtain
\bea
\label{IVXV}
{\cal S}&=&{V_3\over\kappa^2}\left(3\pm{2\over\sqrt{1+{2c\Lambda
\kappa^4\over 3}}}\right)\left(4\pi r_H^3+24\epsilon k\pi r_H\right)\nn
&&+{\cal S}_0\ .
\eea
Therefore when
\be
\label{IVXVI}
c\Lambda>-{5 \over 6\kappa^4}\ ,
\ee
the entropy is always positive for both of solutions corresponding to the
$\pm$ sign. Especially when $c\Lambda>0$, both of solutions expressing
asymptotically AdS and dS spacetime. Then we cannot exclude one of the
solution from the viewpoint of the entropy.

\section{Application to the brane cosmology\label{Sec6}}

We now consider application of the Gauss-Bonnet gravity to the brane
cosmology. Here we consider $Q=0$ case for the simplicity. In the bulk black
hole spacetime in the Gauss-Bonnet gravity, the motion of the brane can be
described by the Friedmann-like equation. One may employ the method developed
in Ref.\cite{NOO1,NOO2,LNO} to derive the Friedmann equation. Then we obtain
\be
\label{FRW1}
H^2={{\cal G}^2\over{\cal H}^2}-{X(a)\over a^2}\equiv f(a)\;,
\ee
where
\bea
\label{G1}
{\cal{G}}&=&4\eta\pm{12X^{1/2}\over Y^{3/2}}\left\{16\epsilon^2\tilde{\mu}^{2}
(4\epsilon-1)^2a^{-3}\right\}\nn
\eta&\equiv&{6\over\kappa_g^2l}\left(12\epsilon-1\right)\;.
\eea
and
\bea
\label{H1}
&&{\cal{H}}=-\frac{48}{\kappa_g^2}\mp{24\over Y^{3/2}}\left(5\DD^2a^{-2}
\right.\nn
&&\ \left.+6\left(\CC\right)^2a^6-{9\left(4\epsilon-1\right)^3
\epsilon{\tilde{\mu}}\over 2\kappa_g^4}a^2\right)\;,\\
\label{XY}
&&X\equiv\frac{k}{2}+\frac{a^2}{4\epsilon l^2}\pm
{Y^{1/2}\kappa_g^2\over 2\epsilon l^2},\nn
&&Y\equiv -2\epsilon{\tilde{\mu}}(4\epsilon -1)
+\frac{(4\epsilon -1)^2}{4\kappa_g^4} a^4\ .
\eea
Here we have defined the rescaled parameters: $\epsilon\equiv{c\kappa_g^2\over
l^2}$ and $\tilde{\mu}\equiv{l^2\mu\over\kappa_g^4}$. The standard
FRW equations for 4-dimensions can be written as
\bea
\label{F1*}
H^2&=&{8\pi G\over 3}\rho -{k\over 2a^2}\ ,\nn
\dot H&=&-4\pi G\left(\rho +p\right)+{k\over 2a^2}\ .
\eea
Here, $\dot{}$ is the derivative with respect to cosmological time.
Then $\rho $ and $p$ are
\bea
\label{or}
\rho&=&{3\over 8\pi G}\left(f(a)+{k\over 2a^2}\right)\\
\label{op}
p&=&-{1\over 8\pi G}\left({k\over 2a^2}+ af'(a)+3f(a)\right),
\eea
where $'$ denotes the derivative with respect to $a$. Similarly, FRW
equations from the bulk dS BH can be constructed (see \cite{LNO}).

In case of $a\to\infty$, we obtain
\bea
\label{FRW2}
\rho&=&{3\over 8\pi G}\left({1\over l^2}(12\epsilon-1)^2
\left(2\pm 3(4\epsilon-1)\right)^{-2}\right.\nn
&&\left.-\frac{1}{4\epsilon l^2}\mp{4\epsilon-1\over 4\epsilon l^2}
\left(1-\epsilon\tilde{\mu}\frac{4\kappa_g^4}{(4\epsilon-1)a^4}\right)
\right)\nn
\label{r3}
p&=&-{1\over 8\pi G}\left({3\over l^2}(12\epsilon-1)^2
\left(2\pm 3(4\epsilon-1)\right)^{-2}\right.\\
&&\left.-\frac{3}{4\epsilon l^2}\mp{3(4\epsilon-1)\over 4\epsilon l^2}
\mp{1\over l^2}\tilde{\mu}\kappa_g^4a^{-4}\right)\ .
\eea
When we choose the upper sign of $\pm$ and $\mp$,
the expressions in (\ref{FRW2}) and (\ref{r3}) are simplified as
\be
\label{FRW3}
\rho={3\over 8\pi G}{\tilde{\mu}\kappa_g^4\over l^{2}a^{4}}\ ,\quad
p={1\over 8\pi G}{\tilde{\mu}\kappa_g^4\over l^{2}a^{4}}\ .
\ee
Then the energy-momentum tensor is traceless ($T^{\mu}_{\ \mu}=\rho -3p=0$.)
and there should be CFT on the brane. On the other hand, if we choose the
lower signs, we find the FRW equations with the cosmological term:
\bea
\label{FRW4}
H^2&=&{8\pi G\over 3}\rho_m-{k\over 2a^2}+{\Lambda\over 3}\ ,\nn
\dot H&=&-4\pi G\left(\rho+p\right)+{k\over 2a^2}\ ,
\eea
We can divide $\rho$ and $p$ into the sums of the contributions
from the matter and those from the cosmological term:
\be
\label{FRW5}
\rho=\rho_m+\rho_0\ ,\quad p=p_m-\rho_0\ ,\quad \rho_0={\Lambda\over 8\pi
G}\ .
\ee
Then we obtain
\bea
\label{FRW6}
&&\rho_0={3\over 8\pi G}\left\{{1\over l^2 }(12\epsilon-1)^2
\left(5-12\epsilon\right)^{-2}\right.\nn
&&\qquad\left.-{1-2\epsilon\over 2\epsilon l^2 }\right\}\ ,\nn
&&\rho_m=-{3\over 8\pi G}{\tilde{\mu}\kappa_g^4\over l^{2}a^{4}}\ ,\quad
p_m=-{1\over 8\pi G}{\tilde{\mu}\kappa_g^4\over l^{2}a^{4}}\ .
\eea
On the other hand, the energy-momentum tensor $T^{m}_{\ \mu\nu}$
of the matter is again traceless,
\be
\label{FRW7}
T^{m\ \mu}_{\ \mu}=\rho_m-3p_m=0\ .
\ee
Then the matter field should be CFT again.

We now consider the limit of $a\to 0$ when
$-2\epsilon{\tilde{\mu}}(4\epsilon-1)>0$. 
\bea
\label{FRW8}
\rho&=&-3p\nn
&=&{3\over 8\pi G}\left\{{2k\over 25}\right.\nn
&&\left.\mp{21\over 25}{\kappa_g^2\over 2\epsilon l^2}(-2\epsilon
{\tilde{\mu}}(4\epsilon-1))^{1/2}\right\}a^{-2}\;.
\eea
Then the energy-momentum tensor is not traceless. Since $\rho$, $p$ $\sim
a^{-2}$, this limit corresponds to the curvature dominant case.
We now rewrite the FRW equation in the following form:
\bea
\label{FRW9}
H^2&=&{8\pi G\over 3}\left(\rho_m-{3\over 8\pi G}{k\over 2a^2}\right)
+{\Lambda\over 3}\nn
&=&{8\pi G\over 3}\tilde{\rho}-{\tilde{k}\over 2a^2}+{\Lambda\over 3}\ .
\eea
Here $\tilde{\rho}$ and $\tilde{k}$ can be regarded as the effective energy
density and effective $k$:
\be
\label{FRW10}
\tilde{\rho}=\rho_m-{3\over 8\pi G}{k\over 2a^2},\quad\tilde{k}=0\ .
\ee
When $k$ is large, $\tilde{\rho}$ behaves as $a^{-2}$.

When $-2\epsilon{\tilde{\mu}}(4\epsilon-1)<0$, $X$ in (\ref{XY}), ${\cal
H}$ in
(\ref{H1}) and ${\cal G}$ in (\ref{G1}) become complex at $a=0$, which
may tell that there is a minimum of $a$ given by $Y=0$:
\be
\label{FRW11}
a=a_{\rm min}\equiv\left({8\epsilon{\tilde{\mu}}\over(4\epsilon-1)}
\right)^{1\over 4}\kappa_g\ .
\ee
When $a=a_{\rm min}$, we have
\bea
\label{FRW12}
\rho&=&-3p\nn
&=&{3\over 8\pi G}\left\{{2k\over 81}\left({2\epsilon{\tilde{\mu}}\over 
(4\epsilon -1)}\right)^{-1\over 2}(2\kappa_g^2)^{-1} \right.\nn
&&\left.-{77\over 81}{1\over 4\epsilon l^2}\right\}\;.
\eea
Then by dividing them into the contributions from the
cosmological term and the matter, we find
\bea
\label{FRW13}
\rho_0&=&-{3\over 8\pi G}{77\over 81}{1\over 4\epsilon l^2}\ ,\nn
\rho_m&=&{3\over 8\pi G}{2k\over 81}\left({2\epsilon{\tilde{\mu}}\over
(4\epsilon-1)}\right)^{-{1\over 2}}(2\kappa_g^2)^{-1}\nn
p_m&=&-{1\over 8\pi G}{2k\over 81}\left({2\epsilon{\tilde{\mu}}\over
(4\epsilon-1)}\right)^{-{1\over 2}}(2\kappa_g^2)^{-1}\;.
\eea

As a special case, we can consider the case of $\epsilon =1/4$. Since  
\be
\label{FRW14}
f(a)=-{k\over 2a^2}\ ,
\ee
and $f(a)=H^2\geq 0$, we find $k\leq 0$. In this case, the energy density and
the pressure vanish: $rho=0$, $p=0$. If one put $\epsilon=1/4-\delta^2$ and
$\delta$ is small, the energy density and the pressure are given by
\bea
\label{FRW15}
\rho&=&-3p\nn
&=&\pm{3\over 8\pi G}{\kappa_g^2\over l^2}
\left\{{2l\over a^3}\left({k\over 2}\right.\right.\nn
&&\left.\left.+{a^2\over l^2}\right)^{1/2}-{7\over a^2}
\right\}(2\tilde{\mu})^{1/2}\delta \;.
\eea
Then in the limit of $a\to 0$, we have $|\rho|\gg |p|$ and $\rho\sim a^{-3}$,
which corresponds to ``dust''. On the other hand, when $a\to \infty$, we find
(if $k\geq 0$) $\rho$, $p\sim a^{-2}$, which tells curvature dominant.
When $k=-2<0$, $a$ has a minimum $a=l$.

More applications to the cosmology are given in \cite{LNO,NOO3}.
Especially the solution describing the bounce universe has been found.

\section{Summary\label{Sec7}}

We have obtained exact solutions in $R^2$-gravity when the action does not
contain the term of the square of the Riemann tensor ($c=0$) and when the
$R^2$-terms are the Gauss-Bonnet combination with the electromagnetic fields.
We have also investigated the thermodynamics of the $R^2$-gravity and we have
found that the entropy often becomes negative. The application to the brane
cosmology is interesting and given in \cite{LNO,NOO3}.

\Acknow
{This research is supported in part by the Ministry of Education, Science, 
Sports and Culture of Japan under the grant number 13135208. The author is 
especially indebted to the discussion with S.D. Odintsov and S. Ogushi. He is 
also very grateful for warm hospitality at GRG11 in Tomsk.}

\end{document}